# An Acetate Bound Cobalt Oxide Catalyst for Water Oxidation: Role of Monovalent Anions and Cations In Lowering Overpotential

Subal Dey, Biswajit Mondal, Abhishek Dey*

Department of Inorganic Chemistry, Indian Association for the Cultivation of Science, Kolkata, India, 700032.

Corresponding Author

icad@iacs.res.in

**Abstract:**

Co(II) dissolved in acetate buffer at pH 7 is found to be a good water oxidation catalyst (WOC) showing electrocatalytic water oxidation current significantly greater than Co(II) in phosphate buffer under the same conditions owing to the higher solubility of the former. When electrodeposited on ITO/FTO electrodes it forms acetate bound cobalt(II)oxide based material (Co-Ac-WOC) showing catalytic water oxidation current density of 0.1 mA/cm$^2$ at 830 mV and 1 mA/cm$^2$ at 1 V in a pH 7 buffer solution. The morphology of Co-Ac-OEC is investigated with AFM, HR-TEM and SEM (at different times and electrodeposition potentials). The chemical composition of Co-Ac-OEC is investigated using XPS, EDX, combustion analysis and ATR-FTIR which indicates that this material has a CoO core with chloride and acetate anions bound to the Co center. Sodium is found to be integrated in the Co-Ac-WOC. The presence of the sodium ions and the chloride ions lowers the onset potential for oxygen evolution reaction (OER) by 240 mV relative to the classic Co-Pi at pH 7. The lower onset potential and higher OER current lowers the exchange current density to $10^{-6.7}$ in Co-Ac-OEC relative to $10^{-8}$-$10^{-10}$ in Co-Pi and its derivatives.

**Introduction:**

The depleting sources of fossil fuels and the growing concern over the detrimental impact of their continual use as the primary fuel source, over a century have catalyzed interest in a clean and renewable energy cycle.[1-3] One dilemma inherent to the traditional renewable sources such as wind or sunlight is the fluctuation of the generated power with varying environmental conditions. Hence, an alternative power source that forms high energy chemical bonds from abundant chemical components is highly desirable, so that energy could be generated on demand from storable fuels.[4,5] A hydrogen based economy heralds such potential.[6,7] $H_2$ can be combined with $O_2$ either in a combustion engine (heat energy) or in fuel cell (electrical energy) generating water. Hydrogen being less abundant in the atmosphere, efficient technologies for splitting water into its gaseous components i.e. - hydrogen and oxygen, is in demand. Electrochemical water splitting contains two half-cells. In an $H^+$ is reduced to $H_2$ $H^+ + e^- \rightarrow ½ H_2$ which in the other $H_2O$ is oxidized to $O_2$ as $2H_2O \rightarrow O_2 + 4e^- + 4H^+$.[3,8,9] Although platinum can catalyze both of these half reactions, it is very expensive. Alternatively, catalysts or composite systems with abundant & cost effective 1st row transition metal which could catalyze these reactions at high turnover rates (TOF) with large turnover numbers (TON), are highly desirable.[10] Robust, cheap, and fast catalyst working at low overpotential for proton reduction to molecular hydrogen has been documented in the recent past.[11,12] On the other hand, a suitable catalyst for the oxygen evolution reaction (OER) is rare. The highest turnover frequency (TOF) reported for HER is 7000 $s^{-1}$ while for OER is 100 $s^{-1}$ using a 1st row transition metal.[12,13] Hence, water oxidation has become the bottleneck of the water splitting process, limiting the overall electrolysis rate.

Over the last few years, tremendous efforts have been devoted to develop several elegant material or molecular based water oxidation catalyst/composites (WOC) by electrochemical or photoelectrochemical means.[14-26] The complexes or oxides of Ru and Ir were investigated as efficient catalyst.[13-15, 18, 19, 27-37] The natural WOC $Mn_4Ca$ cluster in Photosystem II is mimicked with complex heterometallic 1st row transition element.[38-42] But, these synthetic multinuclear manganese clusters are found to be inefficient WOC.[16, 19, 43-46] Very recently, homometallic as well as heterometallic oxides or oxide-nanometarial composites of iron, cobalt and nickel have been established as efficient catalysts for this purpose.[21, 47, 48] OER activity by several molecular complexes of iron and cobalt has been evaluated and found to be inefficient for practical purpose.[49-53] A few years ago, *Nocera et. al.* has introduced a simple but elegant Co-Pi catalyst[54-56] by electrodeposition from Co(II)-salt in phosphate for pursuing water oxidation at pH 7 and when this was incorporated within polyoxometalate (POM) motif by Hill *et. al.*,[57, 58] it was found to be an efficient catalyst satisfying several of the above mentioned features for practical applications.[16, 59-62] Both of these complexes have limited solubility in water and hence their rates are greatly limited under homogeneous conditions. Alternatively several efficient WOCs were obtained from cobalt or nickel based materials obtained from their metal salts as precursors and their catalytic OER activity were reported at various pH.[36, 46, 63-67] These composites show catalytic water oxidation at 300-400 mV overpotentials. Nonetheless several investigations have highlighted the potential of cobalt oxide based materials for generating efficient WOC and these cobalt oxide based materials have been found to be appropriate choices for OER.

Here we show that appropriate choice of buffers and electrolyte can control the chemical composition and morphology of cobalt oxide based materials and greatly enhanced OER activity. Cobalt(II)-salts in neutral Na-acetate buffer in the presence of chloride containing supporting electrolyte (e. g. – NaCl, KCl etc.) shows deposition of acetate bound CoO based (Co-Ac-WOC) composite on ITO or FTO electrodes. The constitution and morphology of this catalyst, investigated using XPS, EDX, combustion, AFM FTIR, HRTEM and SEM, is different from various Co-Pi based catalysts. The Co-Ac-OEC catalyst shows overpotentials of OER of only 20 mV and greater exchange current densities by 2-3 orders of magnitude relative to Co-Pi and its derivatives.

## Results & Discussions

### A. Cyclic Voltammetry of $Co^{II}$ salt in acetate buffer & Comparison with $Co^{II}$ salt in phosphate buffer:

The cyclic voltammogram of $Co^{II}$ in an acetate buffers are performed with glassy carbon (GCE), indium doped tin oxide (ITO) and fluorine doped tin oxide (FTO) in 0.2 N Na-acetate buffer at specific pH and with 0.2 N KCl supporting electrolyte in a normal two compartment electrochemical cell if otherwise not mentioned. The potentials are measured against Ag/AgCl (*saturated* KCl) reference electrode and then corrected and represented against NHE. A 0.5 mM solution of Co-acetate in pH 7 buffer shows a large irreversible current with an onset, (determined from extrapolating the faradic current current slope) of 1.15 V during an anodic scan from 0.2 V to 1.6 V at 50 mVs$^{-1}$ using a GC electrode and reaches to a limiting current density of 8 mA/cm$^2$ at 1.5 V (Fig. 1A green). Such OER activity is not observed for any other 1$^{st}$ row transition metal salts (e.g. – Ni, Mn, Fe, Cu etc.) in the same buffer (Fig. S1A & B). The current density increases linearly with the concentration of the catalyst and a rather high current density of 22 mA/cm$^2$ is achieved in 5 mM $Co^{2+}$ solution at 1.54 V (Fig. 1A deep purple). The linear dependence of catalytic current with $Co^{2+}$ concentration indicates a 1$^{st}$ order rate dependence of $Co^{2+}$ ions in OER. The onset potential and the limiting catalytic current densities obtained at lower concentrations (< 0.5 mM of $Co^{2+}$) for $Co^{II}$ in an acetate buffer is very much comparable to those obtained in phosphate buffer.[54, 68] But, at higher $Co^{2+}$ concentrations the current density is greatly increased and 0.1 mA/cm$^2$ current density is obtained at 1.1 V (onset potential) in an acetate buffer. Such high current densities at these potentials could not be achieved in a phosphate buffer due to the precipitation of $Co_3(PO_4)_2$ from the medium when $Co^{2+}$ concentration reaches ~ 0.5 mM ($K_{sp}$ of $Co_3(PO_4)_2$ = 2 X 10$^{-35}$).

On the other hand when cyclic voltammogram was recorded using ITO and FTO surface, a large current is observed with an onset of 0.91 V and simultaneous formation of gas bubbles on the electrode surface is observed indicating evolution of $O_2$ due to water oxidation. This gas is confirmed to be oxygen by its reduction *in situ* during a cathodic scan by FTO electrode, a method originally introduced by Meyer *et.al.* (Fig. S2).[69] Gradual deposition of a gray material on the ITO or FTO surface is observed during the water oxidation. The onset potential is gradually lowered and the current density is increased in subsequent scans (Fig. S3). It should be mentioned that Co-Pi,[54] $Co^{II}$ deposited on ITO in a phosphate buffer, shows OER activity at considerably higher potentials and with significant lower current densities under the same conditions (Fig. 1B, light purple). The onset potential for the Co-Ac-WOC catalyst is only 0.83 V, *i.e.* the overpotential for OER is only 20 mV, lower than any of the WOCs reported till date. The OER property has also been evaluated with fluorine doped indium oxide (FTO) electrodes and similar onset potential of 0.83 V is attained after the initial scan generated a thin catalytic film of Co-Ac-WOC on the electrode surface (Fig. S4). The onset potential for water oxidation obtained for the Co-Ac-WOC is ~ 240 mV lower than that of the $Co^{2+}$ in acetate buffer solution i.e. as described earlier for GCE (Fig. 1A). The deposited material retains its catalytic activity even in a $Co^{2+}$-free acetate buffer. Further experiments described later are performed on electrodes modified by electrodiposition at 1.2 V for either 3 mintutes or by passing 50 mC/cm$^2$ charge in 5 mM $Co^{2+}$-solution in 0.2 M acetate buffer (pH 7) with 0.2 M potassium chloride as supporting electrolyte.

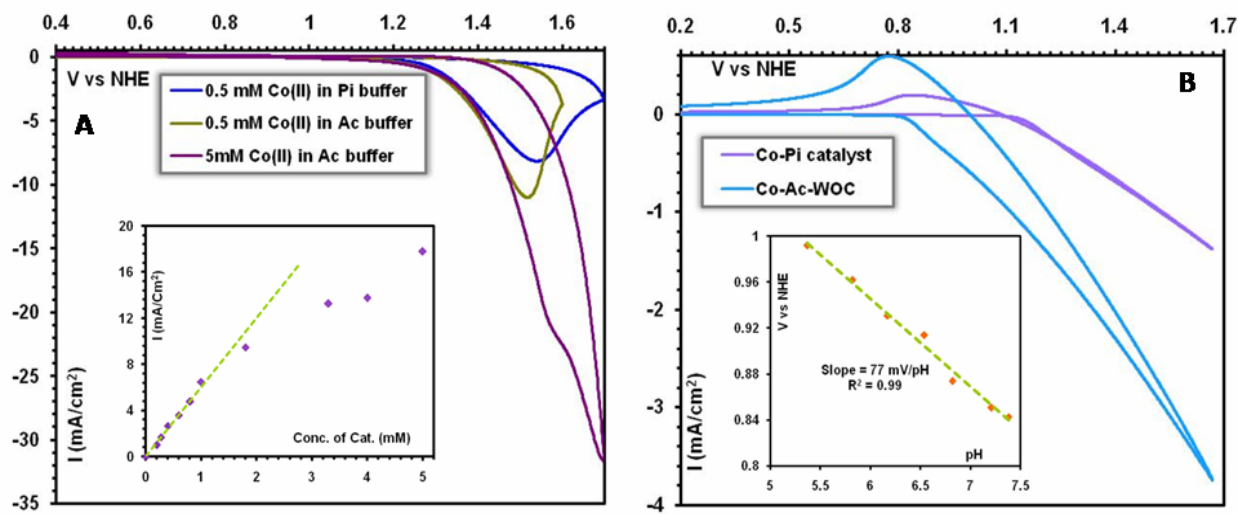

**Figure 1: Co-Pi catalyst, introduced by Nocera *et. al.* vs Co-Ac-OER catalyst.** A) Electrocatalytic OER activity of Co(II) salt in acetate buffer at pH 7 with GC electrode in 0.5 mM and 5 mM concentration and compared with 0.5 mM Co-Pi catalyst. (*Inset*) normalized limiting current density ($I_{cat}$) vs concentration (mM) of catalyst plot. Linear dependence of the current with [$Co^{2+}$] in solution indicates a 1$^{st}$ order reaction kinetics. B) Electrocatalytic OER activity of the electrodeposited Co-Ac-WOC on an ITO electrode at pH 7. (*Inset*) Potential at 0.1 mA/cm$^2$ vs pH plot. 77 mV/pH slop is indicating the involvement of PCET. In both cases Co-Pi has been overlayed for direct comparison.

The OER using Co-Ac-WOC deposited ITO electrodes is investigated in buffers having different pHs (pH 5.4 to pH 7.4). The data indicate that the onset of catalysis is shifted to the higher potentials as the pH is lowered. The onset potential vs pH plot shows a 77mV/pH shift, indicating that a proton coupled electron transfer (PCET) is involved in the OER catalyzed by Co-Ac-WOC as reported for other metal oxide based systems (Fig. 1b *inset*). Specifically Co-Pi shows 64 mV/pH slope in its OER activity.[65, 70-72] Below pH 5 a large positive shift in the onset potential and diminished current density is observed. At higher pHs under the experimental conditions, a black and less active material, likely to be oxides of Co, is deposited on the electrode.

**B. Characterization of Co-Ac-OER deposited on the electrode:**

Previously, these deposits from $Co^{2+}$ solutions have been proposed to be Co(O)OH based materials.[36, 73, 74] The vibrations at 581 $cm^{-1}$, 716$cm^{-1}$ and 761$cm^{-1}$ in the ATR-FTIR spectrum of Co-Ac-OER indicate that it is indeed a cobalt oxide based material (Fig. 2A). The vibration at 1723 $cm^{-1}$ and 1243 $cm^{-1}$ corresponds to the C=O stretch and in plane O-C-O bend of the carboxylate group of the acetate anion. This indicates the presence of $AcO^-$ anion in the material. The C=O stretch at 1723 $cm^{-1}$ and –COO bending mode at 1243 $cm^{-1}$ of the $AcO^-$ in Co-Ac-WOC is lower than those observed in ionic Na-OAc or KOAc complexes. This is characteristic of transition metal bound $AcO^-$ ion and suggests that the $AcO^-$ present in Co-Ac-WOC is likely bound to the cobalt centers in the deposited material. The scanning electron microscopy (SEM) topography of the electrodeposited materials on the ITO surface showed initial formation of an amorphous powder like material (Fig. 2B). Time dependent SEM of the ITO surface shows that the amorphous structure replicates layer by layer, leading to the enlargement of the particle size and formation of stacked plate like structure. SEM images of the ITO surface held at various potentials indicate that the morphology of the material remains unchanged and only the amount of deposited materials increases as the applied potential is increased during electrodiposition. Atomic force microscopy (AFM) images of the modified ITO surfaces indicate that the average heights of these plates are ~ 100 – 120 nm (Fig. 2D & E). High resolution transmission micrograph reproduced the particle size estimates and morphology (Fig. 2C) obtained with SEM.

The energy dispersive X-ray spectroscopy (EDX) analysis enables determination of the elemental composition of a material (Fig. 2F). The data show the presence of cobalt, chloride, sodium and oxygen which is also confirmed by the X-ray photoelectron spectroscopy (XPS) and will be discussed subsequently (Fig. 3). The combination of combustion analysis (estimate of C, H and N by weight) and EDX (estimate of Co, Cl, Na, O by relative abundance) suggests that the relative abundance of these elements Co:O:C:Cl:Na is approximately 24:70.5:8:2:3.5 for the electrodeposited Co-Ac-OEC material. The source of C within the composite is the acetate anion present in the buffer solution. EDX spectroscopy probes deeply into the material and combustion analysis uses bulk of the deposited materials. However catalysis occurs at the surface of these electrodeposited materials and XPS is the better probe of the composition of the catalytically active surface.

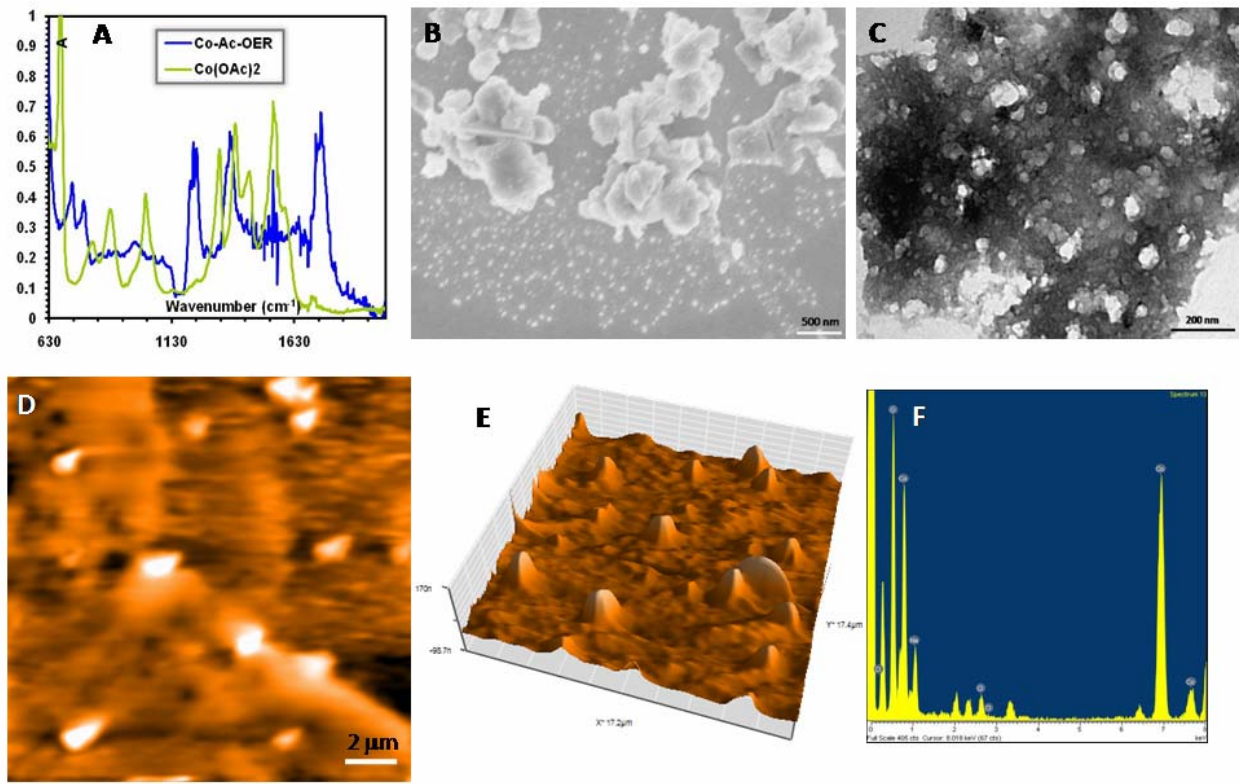

**Figure 2:** Characterization of the Co-Ac-OER. **(A)** IR spectra of Co-Ac-WOC and Co(II)-acetate. (B) Scanning electron microscopy (SEM) topography of the ITO surface with deposited Co-Ac-WOC. (C) Transmisson emission microscope (TEM) image of the Co-Ac-WOC. b & c indicates amorphous composite. (D) AFM topography of the Co-Ac-WOC composite on ITO electrode. (E) 3-dimentional view of the Co-Ac-WOC deposited ITO surface and (F) The EDX spectra of the deposited Co-Ac-WOC. The peak of Cu in EDX spectrum arises from the grid used for TEM.

XPS data show major $2p^{3/2}$ and $2p^{1/2}$ ionization at 781.1 eV and 796. eV of Co, repectively. These ionizations have their corresponding satellite peaks at 786.7 eV and 804.3 eV characteristic of CoO type materials.[75] The presence of CoO type material is also indicated by the Co $3p^{3/2}$ ionization 61.2 eV. There are two overlapping peaks at 530.4 and 532.4 eV are representing 1s ionizations of oxygen atoms from oxide and acetate, respectively.[75, 76] The peaks at 285.1 eV and 288.3 eV represent ionization of carbon 1s core level of the methyl and carboxylate carbon from the acetate respectively.[76] The satellite peak at 293.6 eV is the characteristic of transition metal ligated acetate group and indicates the coordination of the acetate anion to the Co centers present in these materials as initially suggested by the ATR-FTIR data.[77] The presence of chloride and sodium ions in this composite is also confirmed from their 2p and 1s binding energies at 199.7 eV and 1072.3 eV respectively. Note that the chloride 2p ionization at 199.7 eV is higher than its values in ionic complexes (e.g. KCl, NaCl, etc.) and closer to the values reported for complexes where it acts as a bridging ligand between two transition metals.[78] Thus like the acetate the Cl⁻ is likely also bound to the Co centre and most likely as a bridging ligand. Note that even though these electrodiposition is conducted in a solution containing 0.2 M KCl, no potassium ion is detected in the materials. Hence, this material is possibly a form of CoO (FTIR/XPS/EDX), with acetate (FTIR/XPS) and chloride (XPS/EDX) bound to the cobalt centre with some sodium ions (XPS, EDX) dispersed in it. The relative abundance of Co, O, C, Cl and Na, as determined from the XPS data, are 18.3:27.1:5.6:1:3.2. This represents the composition of the catalytic surface as the penetration depth of XPS is only a few nanometers. This is quite different from the composition of the bulk material as obtained from EDX and combustion analysis which shows Co:O:C:Cl:Na ratios of 24:70.5:8:2:3.5 implying that the composition of the material at the surface is different from the core. The higher relative abundance of C on the surface suggests that the AcO⁻ ions (only source of C) are mainly present on the surface of the material. The higher abundance of oxygen in the core and the characteristic Co $2p^{3/2}$, $2p^{1/2}$, $3p^{3/2}$ ionization values is consistent with the presence of a CoO core. The heterogeneous composition of the Co-Ac-WOC material, with the anion (in this case AcO⁻) mainly localized on the suface is similar to those reported for the Co-Pi which is obtained from electrodiposition of $Co^{2+}$ in $PO_4^{3-}$ buffer.[56, 79] The XPS study of the deposited materials from the acetate buffer with only potassium and nitrate instead of sodium and chloride showed that neither nitrate nor potassium has been incorporated within these materials. Rather the electrodeposited material is characterized to be oxides of Co(II) with bound acetate. The morphology of this material is changes as well determined by SEM topography (*SI*). In the following section, the effect of the Cl⁻, AcO⁻ and Na⁺ ions an OER is discussed.

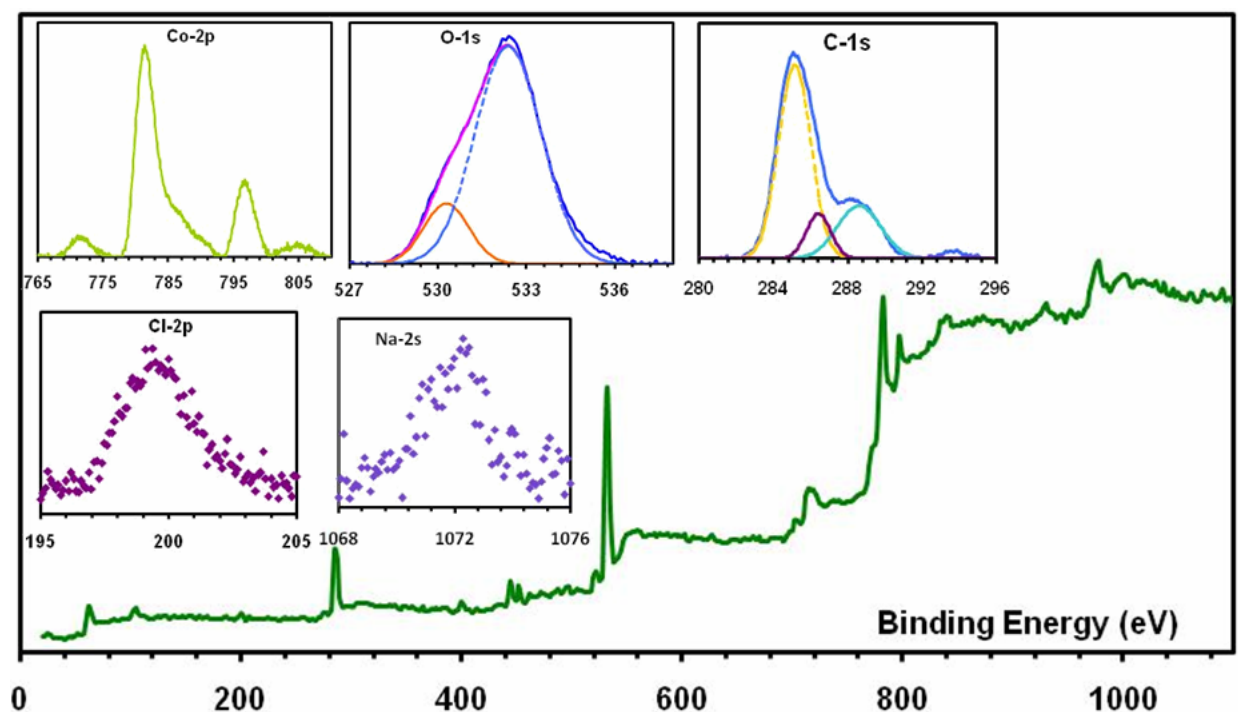

**Figure 3:** XPS characterization of the Co-Ac-WOC. (*inset*) The binding energies of the corresponding elements present (Co, O, C, Cl, Na respectively) in Co-Ac-WOC are indicated.

### C. Role of electrolytes in OER:

To investigate the choice of cobalt salt in sodium acetate buffer with potassium chloride supporting electrolyte, OER is investigated by changing the counter-ion of the Co(II)-salt, electrolyte and buffer compositions at pH 7. For each of these experiments, an ITO electrode is initially modified by electrodiposition from the 5 mM $Co^{2+}$ solution in a specific buffer and electrolytic solution. Then the electrochemical OER is evaluated for this electrodeposited material in the same buffer and electrolytic solution in the absence of dissolved Co(II). In these experiments acetate buffer are used to obtain high solubility to Co (II) ions at pH 7 as discussed earlier. As long as an acetate buffer with NaCl or KCl electrolyte is used; the catalytic activity does not depend on the source of $Co^{2+}$ as any water soluble $Co^{2+}$ salt could be used (*e.g.* – $CoCl_2.6H_2O$, $Co(OAc)_2.4H_2O$, $Co(NO_3)_2.4H_2O$ etc.) (*SI*). On deposition from a solution containing neither chloride nor sodium (i.e. in a 5 mM $Co(NO_3)_2$ solution in a KOAc buffer with $KNO_3$ electrolyte) the modified ITO electrode shows catalytic onset at 1.04 V at pH 7 (Fig. 4, orange trace), which is 220 mV more positive than the value obtained for Co-Ac-WOC in Na-OAc buffer with KCl or NaCl electrolyte. In a Cl$^-$ free buffer containing either KOAc or NaOAc with NaNO$_3$ or KNO$_3$ electrolyte, respectively, the onset potential is ~ 100 mV higher than those obtained with KOAc/NaCl (Fig. 4, cyan), NaOAc/KCl (Fig. 1b, sky blue) or NaOAc/NaCl (Fig. 4, light green). The lowering of catalytic onset potential by Cl$^-$, which is likely bound to the Co-centre as indicated by the XPS data, is likely due to stabilization of highly oxidized species formed during OER by an anionic ligand. Monovalent ions like Na$^+$ has been known to enhance catalytic OER both in natural (Photosystem-II) and synthetic systems.[80] It is important to note that K$^+$ ions present in the electrolytic solution could not be detected in the electrodeposited material. This is likely due to larger size of the K$^+$ ion which does not allow its incorporation in the material. Thus the small amount of Na$^+$ and Cl$^-$ present in Co-Ac-WOC heavily influence its reactivity by lowering down the onset potential by 240 mV.

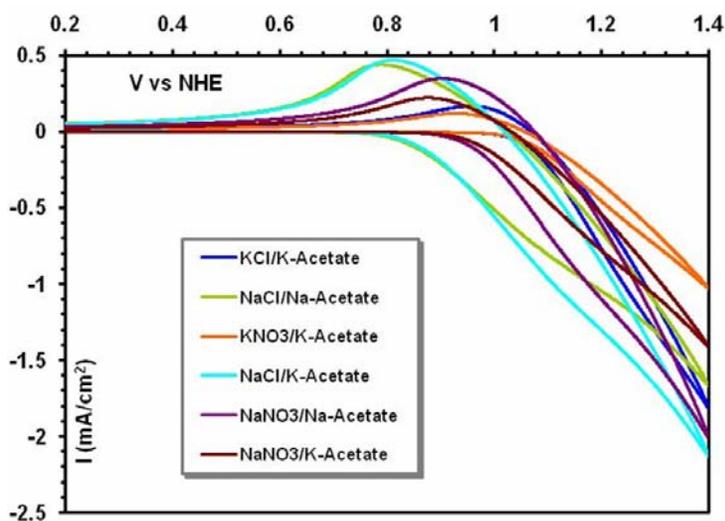

**Figure 4:** Effect of coordinating ligand, buffer & supporting electrolyte. Cyclic voltamograms in solution having various electrolyte compositions.

### D. Bulk electrolysis & Kinetic Rate:

Electrocatalytic water oxidation by Co(II) in acetate buffer indicate that this system shows efficient OER activity under both homogeneous conditions (with GC working electrodes), and the Co-Ac-WOC system shows efficient OER under heterogeneous conditions (with ITO & FTO electrodes). The deposited material on the ITO surface is able to produce oxygen in the presence of small amount (*ca* 0.3 mM) of the $Co^{2+}$-salt in solution over a long period of time and shows a constant catalytic current of 2.5 mA/cm² in a controlled potential electrolysis (CPE) experiment performed at 1.3 V. The faradic efficiency (FY) is determined to be 94%, by measuring the volume of oxygen produced (1.9 ml) by displacement of water at 1 atm pressure (Fig. 5A).[81] The TOF is determined to be 17.8 s$^{-1}$ from the steady current density in CPE and double layer capacitance ($C_{dl}$ = 35 µF/cm²) obtained from the CV at a region which does not show catalytic current. This value is the highest among the any of the oxide based catalyst reported so far.[57, 69] The CPE with a FTO electrode at 1.2 V shows activity very similar to the ITO electrodes (Fig. *SI*). The high solubility of the molecular oxygen in water may be responsible for the lowering of FY from 100% as observed.

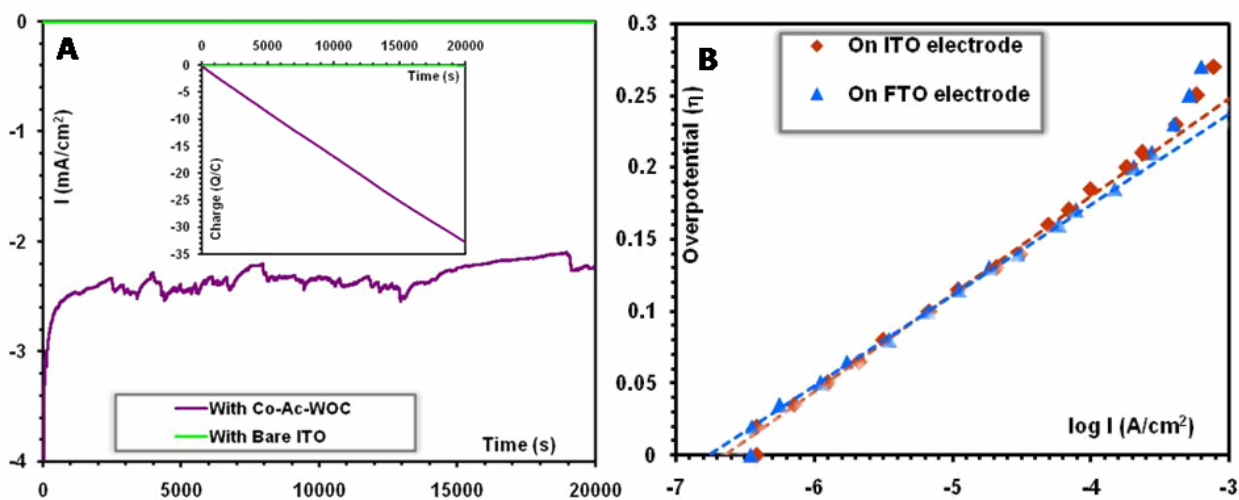

Figure 5: Kinetic activity of the Co-AC-OER. (a) The controlled potential electrolysis at 1.3 V (overpotential 490 mV) with Co-Ac-OER deposited ITO electrode in 0.5 mM Co2+ solution at pH 7. (b) Tafel plot of the Co-Ac-OER at pH 7. The material deposited on ITO shows tafel slop 68 mV/dec with an exchange current density 10$^{-6.6}$ A/cm² and on FTO shows 62 mV/dec with an exchange current density 10$^{-6.7}$ A/cm².

The exchange current density ($J_0$) is the direct measure of the inherent catalytic efficiency of a particular catalyst for a defined electrochemical reaction. To determine the $J_0$, Co-Ac-WOC deposited electrode is prepared from a 5 mM Co-salt solution in acetate buffer by passing 50 mC charge in each experiment followed by thorough washing with fresh water. Bulk electrolysis is performed with increasing overpotentials in an acetate buffer at pH 7 with 0.3 mM $Co^{2+}$ ion present in solution. At each potential stable current density is recorded after 2 hours of pre-electrolysis period allowing the system to attain equilibrium. The log I vs overpotential ($\eta$) plot show that tafel slopes in acetate buffer (pH 7) are 68 mV/decade for ITO electrodes and 62 mV/dec for FTO electrodes (Fig. 5B). These values are comparable to the other 1[st] row transition metal oxide based catalysts of e.g. – Co-Pi catalyst, Ni-Bi catalyst etc.[36, 65] The data show that the catalysis initiates at only 20 mV overpotentials. The exchange current density ($J_0$) for Co-Ac-WOC is determined to be $10^{-6.7}$ when deposited on ITO electrode and $10^{-6.75}$ A/cm$^2$ when deposited on FTO electrode. $J_0$ values for Co-Pi, or any catalyst with small modification of the Co-Pi, are in the range of $10^{-8} - 10^{-10}$ A/cm$^2$.[36] The $J_0$ value of the metallic Pt is reported to be $10^{-6}$ A/cm$^2$.[82] Very recently, some mixed metal oxides have shown to have $J_0$ values at the range of $10^{-7} - 10^{-8}$ A/cm$^2$.[47] Thus the Co-Ac-OEC catalyst has one of the highest $J_0$ for any WOCs reported till date.

**WOC Activity with Silicon-Carbide electrode:**

The cyclic voltammogram has also been performed with freshly cleaned silicon carbide electrodes (used in domestic $MnO_2/Zn/NH_4Cl$ dry batteries), The data indicate OER with an onset at 1.05 V. The large porous surface of the domestic electrode helped to attain catalytic current densities as high as 9 mA/cm$^2$ at only 385 mV overpotential. The CPE with domestic graphite electrode at 1.2 V in 5 mM $Co^{2+}$ solution results in accumulation of 77 C charge through the electrode with a constant 9 mA/cm$^2$ current density over 2 hours and simultaneously 4.5 ml of oxygen is accumulated. The FY for this set up was calculated to be 93% (*SI*).

**Conclusion**

Here, homogeneous and heterogeneous electrocatalytic OER activity of Co(II) in acetate buffer is investigated in and found to be efficient for practical purposes. The higher solubility of Co(II) in acetate buffer leads to enhance homogeneous OER activity relative to the Co-OEC in phosphate buffer. Acetate and chloride bound containing sodium ion CoO based material is electrodeposited on ITO & FTO electrode. This Co-Ac-WOC modified electrode shows high OER activity under heterogeneous condition with only 20 mV overpotentials. These composite electrodes show the highest exchange current density ($\sim 10^{-6.7}$ A/cm$^2$) among the any of the 1$^{st}$ row transition metal WOCs reported so far ($\sim 10^{8-11}$ A/cm$^2$) and even more than the rutile (RuO$_2$) or iridium oxides (IrO$_x$) ($\sim 10^{-(8-9)}$ A/cm$^2$).[37] This material is stable under prolonged electrolysis where it produces current density as high as 2.5 mA/cm$^2$ with only 485 mV overpotential. The cobalt bound acetate and chloride as ions and intervening sodium ions present in the material contribute to the enhancement of OER activity relative to the classic Co-Pi. This catalyst show highest current density ($\sim$ 9 mA/cm$^2$) at only 385 mV overpotential with cheap Si-C electrode. Further detailed structural, mechanistic and spectroscopic investigations with Co-Ac-WOC are undergoing, which may allow further improvement in the OER catalysis.

*Supporting Information*

Further detailed electrochemical results are given in Figures S1 to S10 as well as the additional SEM micrographs are available in *SI*. The double layer capacitance measurements have been explained in *SI*. This material is available free of charge via the Internet at http://pubs.acs.org.

*Acknowledgements*

This research is funded by BRNS grant number 2011/36/12-BRNS/, DAE, India. S.D. acknowledges a CSIR-SRF fellowship. B. M. acknowledges CSIR-SPM-JRF fellowship. The IACS XPS facility funded by DST unit of Nano-science is gratefully acknowledged.

**Reference**

1.      Chow, J.; Kopp, R. J.; Portney, P. R., Science **2003,** 302, 1528.
2.      Walter, M. G.; Warren, E. L.; McKone, J. R.; Boettcher, S. W.; Mi, Q.; Santori, E. A.; Lewis, N. S., Chemical Reviews **2010,** 110, 6446.
3.      Cook, T. R.; Dogutan, D. K.; Reece, S. Y.; Surendranath, Y.; Teets, T. S.; Nocera, D. G., Chemical Reviews **2010,** 110, 6474.
4.      Lewis, N. S.; Nocera, D. G., Proceedings of the National Academy of Sciences **2006,** 103, 15729.

5.	Gray, H. B., Nat Chem **2009,** 1, 7.
6.	Armstrong, F. A.; Belsey, N. A.; Cracknell, J. A.; Goldet, G.; Parkin, A.; Reisner, E.; Vincent, K. A.; Wait, A. F., Chemical Society Reviews **2009,** 38, 36.
7.	Armstrong, F. A.; Hirst, J., Proceedings of the National Academy of Sciences **2011,** 108, 14049.
8.	Nocera, D. G., Accounts of Chemical Research **2012,** 45, 767.
9.	Meyer, T. J., Nat Chem **2012,** 3, 757.
10.	Du, P.; Eisenberg, R., Energy & Environmental Science **2012,** 5, 6012.
11.	Mondal, B.; Sengupta, K.; Rana, A.; Mahammed, A.; Botoshansky, M.; Dey, S. G.; Gross, Z.; Dey, A., Inorganic Chemistry **2013,** 52, 3381.
12.	Dey, S.; Rana, A.; Dey, S. G.; Dey, A., ACS Catalysis **2013,** 3, 429.
13.	Barnett, S. M.; Goldberg, K. I.; Mayer, J. M., Nat Chem **2012,** 4, 498.
14.	Polyansky, D. E.; Muckerman, J. T.; Rochford, J.; Zong, R.; Thummel, R. P.; Fujita, E., Journal of the American Chemical Society **2011,** 133, 14649.
15.	Schley, N. D.; Blakemore, J. D.; Subbaiyan, N. K.; Incarvito, C. D.; D'Souza, F.; Crabtree, R. H.; Brudvig, G. W., Journal of the American Chemical Society **2011,** 133, 10473.
16.	Risch, M.; Klingan, K.; Ringleb, F.; Chernev, P.; Zaharieva, I.; Fischer, A.; Dau, H., ChemSusChem **2012,** 5, 542.
17.	Gerken, J. B.; Chen, J. Y. C.; Massé, R. C.; Powell, A. B.; Stahl, S. S., Angewandte Chemie International Edition **2012,** 51, 6676.
18.	Hintermair, U.; Hashmi, S. M.; Elimelech, M.; Crabtree, R. H., Journal of the American Chemical Society **2012,** 134, 9785.
19.	Busch, M.; Ahlberg, E.; Panas, I., The Journal of Physical Chemistry C **2013,** 117, 288.
20.	Singh, A.; Spiccia, L., Coordination Chemistry Reviews **2013,** 257, 2607.
21.	Louie, M. W.; Bell, A. T., Journal of the American Chemical Society **2013,** 135, 12329.
22.	Hintermair, U.; Sheehan, S. W.; Parent, A. R.; Ess, D. H.; Richens, D. T.; Vaccaro, P. H.; Brudvig, G. W.; Crabtree, R. H., Journal of the American Chemical Society **2013,** 135, 10837.
23.	Swierk, J. R.; Mallouk, T. E., Chem Soc Rev **2013,** 42, 2357.
24.	Young, K. J.; Martini, L. A.; Milot, R. L.; Snoeberger Iii, R. C.; Batista, V. S.; Schmuttenmaer, C. A.; Crabtree, R. H.; Brudvig, G. W., Coordination Chemistry Reviews **2012,** 256, 2503.
25.	Tran, P. D.; Artero, V.; Fontecave, M., Energy & Environmental Science **2010,** 3, 727.
26.	Artero, V.; Chavarot-Kerlidou, M.; Fontecave, M., Angewandte Chemie International Edition **2011,** 50, 7238.
27.	Gersten, S. W.; Samuels, G. J.; Meyer, T. J., Journal of the American Chemical Society **1982,** 104, 4029.
28.	Duan, L.; Araujo, C. M.; Ahlquist, M. r. S. G.; Sun, L., Proceedings of the National Academy of Sciences **2012,** 109, 15584.
29.	Kaveevivitchai, N.; Chitta, R.; Zong, R.; El Ojaimi, M.; Thummel, R. P., Journal of the American Chemical Society **2012,** 134, 10721.
30.	Grotjahn, D. B.; Brown, D. B.; Martin, J. K.; Marelius, D. C.; Abadjian, M.-C.; Tran, H. N.; Kalyuzhny, G.; Vecchio, K. S.; Specht, Z. G.; Cortes-Llamas, S. A.; Miranda-Soto, V.; van Niekerk, C.; Moore, C. E.; Rheingold, A. L., Journal of the American Chemical Society **2011,** 133, 19024.
31.	Planas, N.; Christian, G.; Roeser, S.; Mas-Marzá, E.; Kollipara, M.-R.; Benet-Buchholz, J.; Maseras, F.; Llobet, A., Inorganic Chemistry **2012,** 51, 1889.
32.	Hull, J. F.; Balcells, D.; Blakemore, J. D.; Incarvito, C. D.; Eisenstein, O.; Brudvig, G. W.; Crabtree, R. H., Journal of the American Chemical Society **2009,** 131, 8730.
33.	Romain, S.; Vigara, L.; Llobet, A., Accounts of Chemical Research **2009,** 42, 1944.
34.	Zhang, Y.; Ren, T., Chemical Communications **2012,** 48, 11005.

35. Nakagawa, T.; Bjorge, N. S.; Murray, R. W., *Journal of the American Chemical Society* **2009,** *131,* 15578.
36. Gerken, J. B.; McAlpin, J. G.; Chen, J. Y. C.; Rigsby, M. L.; Casey, W. H.; Britt, R. D.; Stahl, S. S., *Journal of the American Chemical Society* **2011,** *133,* 14431.
37. Burke, L. D.; Murphy, O. J.; O'Neill, J. F.; Venkatesan, S., *Journal of the Chemical Society, Faraday Transactions 1: Physical Chemistry in Condensed Phases* **1977,** *73,* 1659.
38. Ferreira, K. N.; Iverson, T. M.; Maghlaoui, K.; Barber, J.; Iwata, S., *Science* **2004,** *303,* 1831.
39. Yano, J.; Kern, J.; Sauer, K.; Latimer, M. J.; Pushkar, Y.; Biesiadka, J.; Loll, B.; Saenger, W.; Messinger, J.; Zouni, A.; Yachandra, V. K., *Science* **2006,** *314,* 821.
40. Dismukes, G. C.; Brimblecombe, R.; Felton, G. A. N.; Pryadun, R. S.; Sheats, J. E.; Spiccia, L.; Swiegers, G. F., *Accounts of Chemical Research* **2009,** *42,* 1935.
41. Brimblecombe, R.; Koo, A.; Dismukes, G. C.; Swiegers, G. F.; Spiccia, L., *Journal of the American Chemical Society* **2010,** *132,* 2892.
42. Mukherjee, S.; Stull, J. A.; Yano, J.; Stamatatos, T. C.; Pringouri, K.; Stich, T. A.; Abboud, K. A.; Britt, R. D.; Yachandra, V. K.; Christou, G., *Proceedings of the National Academy of Sciences* **2012**.
43. Sartorel, A.; Carraro, M.; Toma, F. M.; Prato, M.; Bonchio, M., *Energy & Environmental Science* **2012,** *5,* 5592.
44. Wiechen, M.; Zaharieva, I.; Dau, H.; Kurz, P., *Chemical Science* **2012,** *3,* 2330.
45. Toma, F. M.; Sartorel, A.; Iurlo, M.; Carraro, M.; Parisse, P.; Maccato, C.; Rapino, S.; Gonzalez, B. R.; Amenitsch, H.; Da Ros, T.; Casalis, L.; Goldoni, A.; Marcaccio, M.; Scorrano, G.; Scoles, G.; Paolucci, F.; Prato, M.; Bonchio, M., *Nat Chem* **2010,** *2,* 826.
46. Jiao, F.; Frei, H., *Energy & Environmental Science* **2010,** *3,* 1018.
47. Smith, R. D. L.; Prévot, M. S.; Fagan, R. D.; Zhang, Z.; Sedach, P. A.; Siu, M. K. J.; Trudel, S.; Berlinguette, C. P., *Science* **2013,** *340,* 60.
48. Smith, R. D. L.; Prévot, M. S.; Fagan, R. D.; Trudel, S.; Berlinguette, C. P., *Journal of the American Chemical Society* **2013,** *135,* 11580.
49. Fillol, J. L.; Codolà, Z.; Garcia-Bosch, I.; Gómez, L.; Pla, J. J.; Costas, M., *Nat Chem* **2011,** *3,* 807.
50. Rigsby, M. L.; Mandal, S.; Nam, W.; Spencer, L. C.; Llobet, A.; Stahl, S. S., *Chemical Science* **2012,** *3,* 3058.
51. Wasylenko, D. J.; Palmer, R. D.; Berlinguette, C. P., *Chemical Communications* **2013,** *49,* 218.
52. Ellis, W. C.; McDaniel, N. D.; Bernhard, S.; Collins, T. J., *Journal of the American Chemical Society* **2010,** *132,* 10990.
53. Dogutan, D. K.; McGuire, R.; Nocera, D. G., *Journal of the American Chemical Society* **2011,** *133,* 9178.
54. Kanan, M. W.; Nocera, D. G., *Science* **2008,** *321,* 1072.
55. Lutterman, D. A.; Surendranath, Y.; Nocera, D. G., *Journal of the American Chemical Society* **2009,** *131,* 3838.
56. Kanan, M. W.; Surendranath, Y.; Nocera, D. G., *Chemical Society Reviews* **2009,** *38,* 109.
57. Yin, Q.; Tan, J. M.; Besson, C.; Geletii, Y. V.; Musaev, D. G.; Kuznetsov, A. E.; Luo, Z.; Hardcastle, K. I.; Hill, C. L., *Science* **2010,** *328,* 342.
58. Stracke, J. J.; Finke, R. G., *Journal of the American Chemical Society* **2011,** *133,* 14872.
59. Young, E. R.; Nocera, D. G.; Bulovic, V., *Energy & Environmental Science* **2010,** *3,* 1726.
60. Zhong, D. K.; Gamelin, D. R., *Journal of the American Chemical Society* **2010,** *132,* 4202.
61. Reece, S. Y.; Hamel, J. A.; Sung, K.; Jarvi, T. D.; Esswein, A. J.; Pijpers, J. J. H.; Nocera, D. G., *Science* **2011,** *334,* 645.
62. Lai, Y.-H.; Lin, C.-Y.; Lv, Y.; King, T. C.; Steiner, A.; Muresan, N. M.; Gan, L.; Wright, D. S.; Reisner, E., *Chemical Communications* **2013,** *49,* 4331.
63. Ahn, H. S.; Tilley, T. D., *Advanced Functional Materials* **2013,** *23,* 227.

64. Zhu, G.; Glass, E. N.; Zhao, C.; Lv, H.; Vickers, J. W.; Geletii, Y. V.; Musaev, D. G.; Song, J.; Hill, C. L., Dalton Transactions **2012,** 41, 13043.
65. Dincǎ, M.; Surendranath, Y.; Nocera, D. G., Proceedings of the National Academy of Sciences **2010,** 107, 10337.
66. Esswein, A. J.; McMurdo, M. J.; Ross, P. N.; Bell, A. T.; Tilley, T. D., The Journal of Physical Chemistry C **2009,** 113, 15068.
67. McCool, N. S.; Robinson, D. M.; Sheats, J. E.; Dismukes, G. C., Journal of the American Chemical Society **2011,** 133, 11446.
68. Surendranath, Y.; DincaÌŒ, M.; Nocera, D. G., Journal of the American Chemical Society **2009,** 131, 2615.
69. Kent, C. A.; Concepcion, J. J.; Dares, C. J.; Torelli, D. A.; Rieth, A. J.; Miller, A. S.; Hoertz, P. G.; Meyer, T. J., Journal of the American Chemical Society **2013,** 135, 8432.
70. Surendranath, Y.; Kanan, M. W.; Nocera, D. G., Journal of the American Chemical Society **2010,** 132, 16501.
71. Esswein, A. J.; Surendranath, Y.; Reece, S. Y.; Nocera, D. G., Energy & Environmental Science **2011,** 4, 499.
72. Chen, Z.; Vannucci, A. K.; Concepcion, J. J.; Jurss, J. W.; Meyer, T. J., Proceedings of the National Academy of Sciences **2011,** 108, E1461.
73. Bockris, J. O.; Otagawa, T., The Journal of Physical Chemistry **1983,** 87, 2960.
74. Bockris, J. O. M.; Otagawa, T., Journal of The Electrochemical Society **1984,** 131, 290.
75. McIntyre, N. S.; Cook, M. G., Analytical Chemistry **1975,** 47, 2208.
76. Dennis, A. M.; Howard, R. A.; Kadish, K. M.; Bear, J. L.; Brace, J.; Winograd, N., Inorganica Chimica Acta **1980,** 44, L139.
77. Bozack, M. J.; Zhou, Y.; Worley, S. D., The Journal of Chemical Physics **1994,** 100, 8392.
78. Wu, H.-M.; Chen, S.-A., Synthetic Metals **1988,** 26, 225.
79. Kanan, M. W.; Yano, J.; Surendranath, Y.; Dincǎ, M.; Yachandra, V. K.; Nocera, D. G., Journal of the American Chemical Society **2010,** 132, 13692.
80. Tsui, E. Y.; Tran, R.; Yano, J.; Agapie, T., Nat Chem **2013,** 5, 293.
81. The catalytic activity of Co-Ac-WOC is gradually decreased during CPE experiments in a buffer without any $Co^{2+}$. This is likely due to loss of $Co^{2+}$ from the material. However, steady activity can be maintained in the presence of 0.3 mM $Co^{2+}$ in solution.
82. Markovic , N. M., In Handbook of Fuel Cells – Fundamentals, Technology and Applications, W. Vielstich, A. L., and H. A. Gasteiger, Ed. John Wiley & Sons Ltd.: New York, 2003; Vol. 2, p 374.

TOC

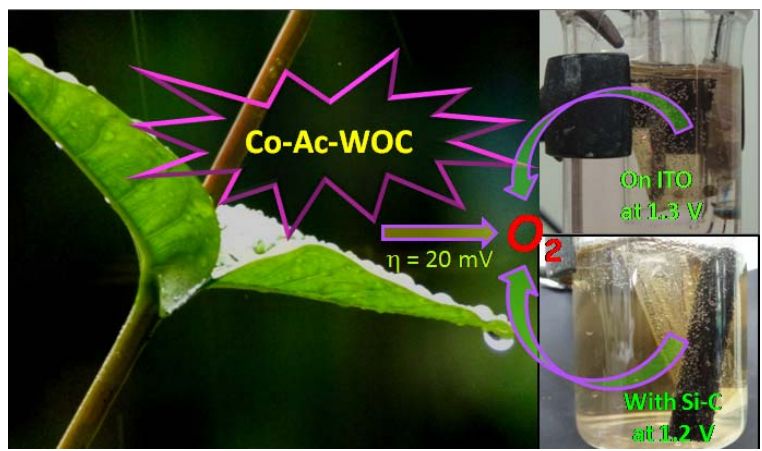

# An Acetate Bound Cobalt Oxide Catalyst for Water Oxidation: Role of Monovalent Anions and Cations In Lowering Overpotential

*Subal Dey, Biswajit Mondal, Abhishek Dey*[*]

*Department of Inorganic Chemistry, Indian Association for the Cultivation of Science, Kolkata, India, 700032.*

*Corresponding Author : icad@iacs.res.in*

**General Procedures**

All the starting materials are purchased from Merck Chemicals Ltd, India through vendors and are used without further puridications. The indium doped tin oxide (ITO) and fluorine doped tin oxide electrodes are purchased from Sigma-Aldrich India and are used after cleaning. The slicon-carbide electrode was directly extracted from market available batteries and washed thoroughly with concentrated $HNO_3$ and water and dried for overnight for further use. For all the experiments triple distilled water is used. The electrochemical measurements have been performed with Pine CHI 720D bipotentiostate using a Pt wire counter electrodes (CHI 115) and Ag/AgCl (sat. KCl) reference ( CHI 111) electrodes. FTIR spectra is recorded in Perkin-Elmer Frontier model using ATR technique. Field-emission scanning electron microscopy (FESEM) was performed with model JEOL JSM-6700F) and high-resolution transmission electron microscopic (HRTEM) topographs are collected from a JEOL-JEM 2010 microscope using 200k eV electron source. Energy-dispersive X-ray spectroscopy (EDS) was done using a HRTEM JEOL-JEM 2010 microscope. X-ray photoelectron spectroscopy (XPS, Omicron, model: 1712-62-11) measurement was done using an Al-Ka radiation source under 15 kV voltage and 5 mA current condition.

Figure S 1 : Cyclic Voltamogram of (A) 5 mM solution of the transition metal salts with GC electrode and (B) with ITO electrode modified with the Ni-Ac-WOC, Fe-Ac-WOC, Mn-Ac-WOC and comparison with Co-Ac-WOC at 50 mVs$^{-1}$ scan rate in acetate buffer (pH 7).

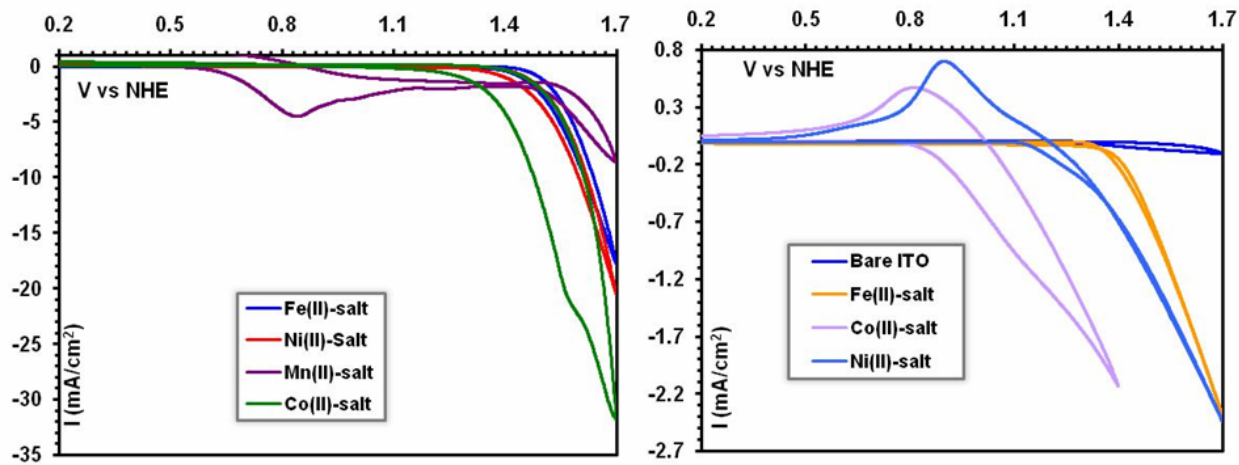

Figure S2: Cyclic voltamogram of the FTO electrod with Co-Ac-WOC in deoxygenated Co$^{2+}$ free buffer (Acetate pH 7) at 50 mVs$^{-1}$. Red arrow indicates the direction of the initial scan. OER & ORR stand for oxygen evolution reaction and ORR for oxygen reduction reaction.

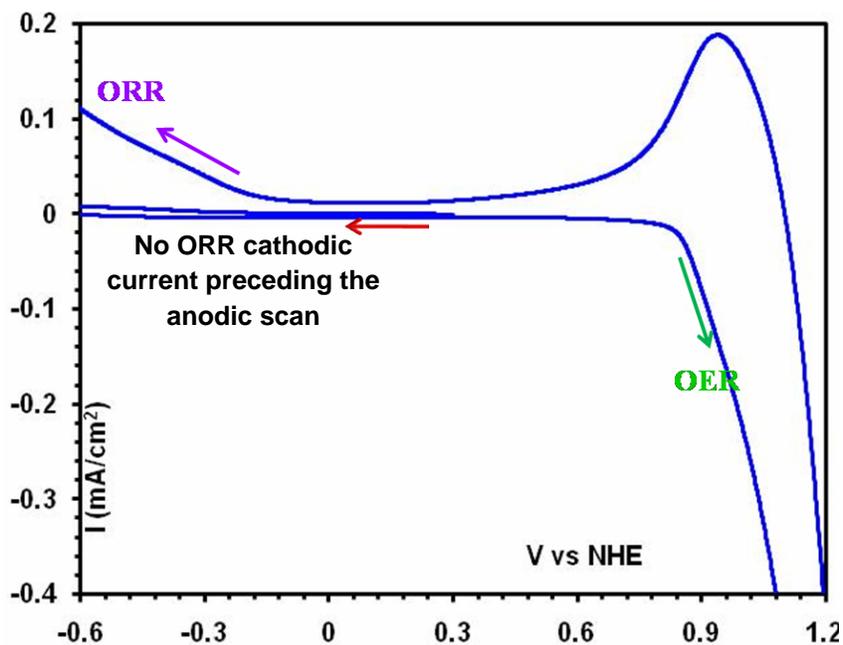

Figure S3: CV with ITO electrode in 1 mM Co$^{2+}$ solution in acetate buffer with KCl electrolyte at pH 7 at 50 mVs$^{-1}$.

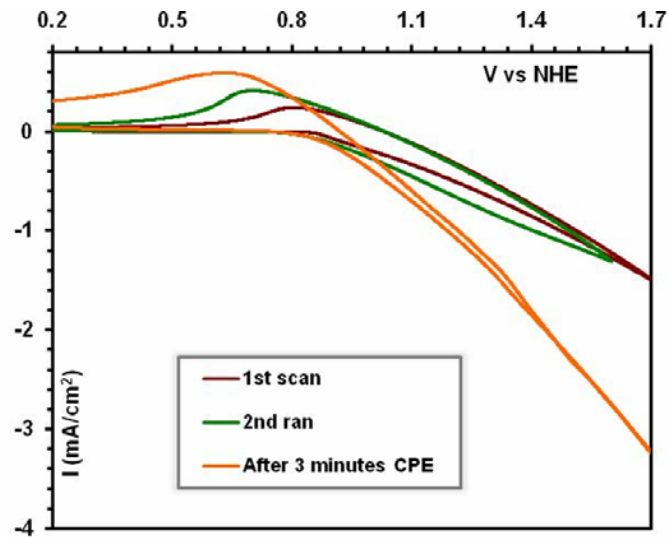

Figure S4: Cyclic Voltamogram with FTO electrode at 50 mVs$^{-1}$ with Co-Ac-WOC at pH 7 acetate buffer and overlayed with the bare FTO electrode at a Co$^{2+}$-free solution under same conditions.

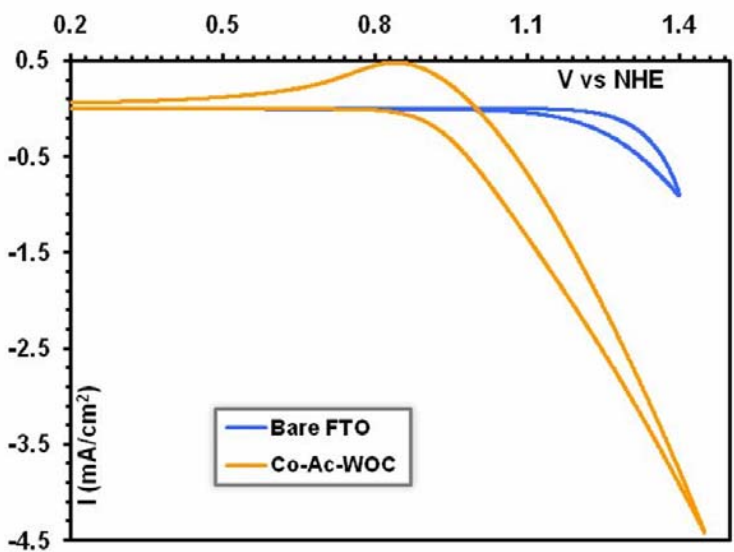

Figure S5: CPE with FTO electrode in pH 7 acetate buffer at 1.2 V vs NHE.

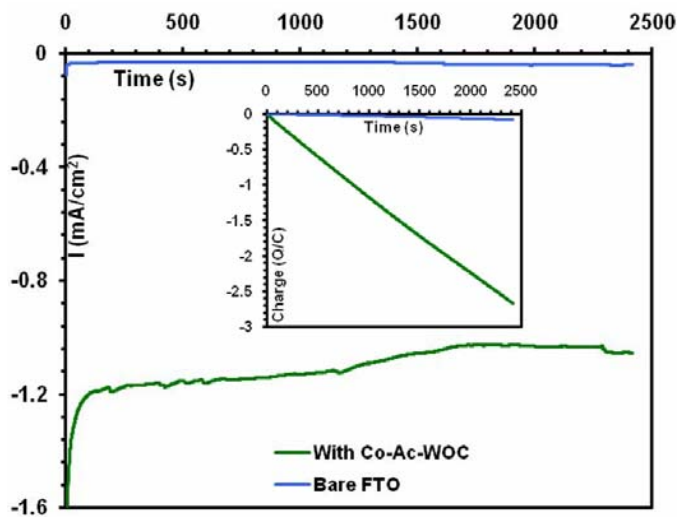

Figure S6: Electrocatalytic water oxidation with GC electrode and 1 mM $Co^{2+}$ solution at pH 7 (acetate buffer). (*Inset*) Limiting current vs (scan rate)$^{1/2}$ plot.

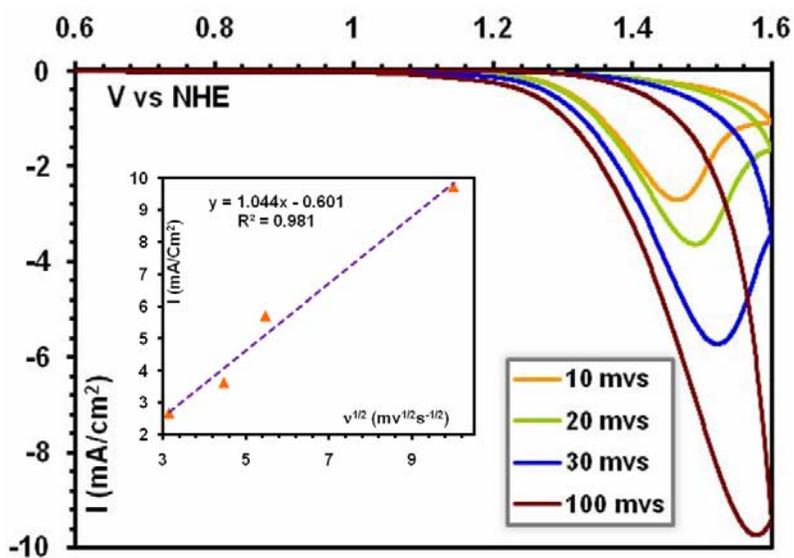

Figure S7 : CV in acetate buffer (pH 7) of with different precursor $Co^{2+}$ salts at 50 $mVs^{-1}$ and ITO electrode.

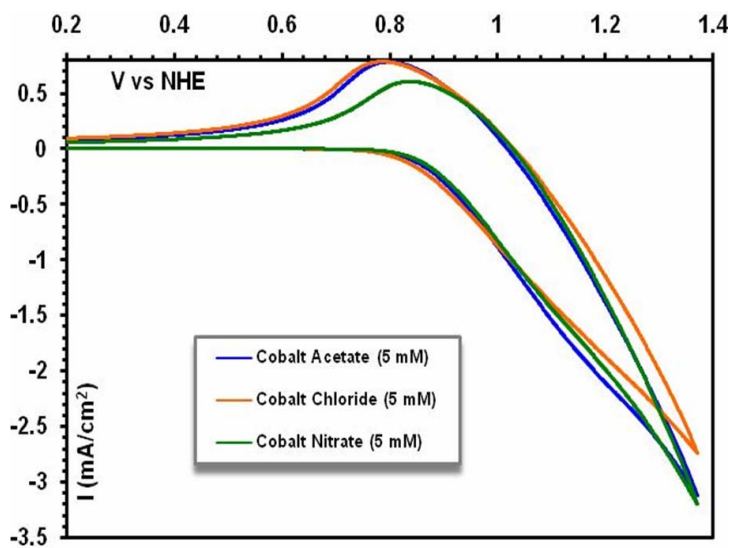

Figure S8: A) Linear Sweep Voltametry (LSV) with Si-C electrodes at pH 7 and 50 $mVs^{-1}$. (B) CPE at 1.2 V vs NHE in acetate buffer (pH 7) with Si-C electrode.

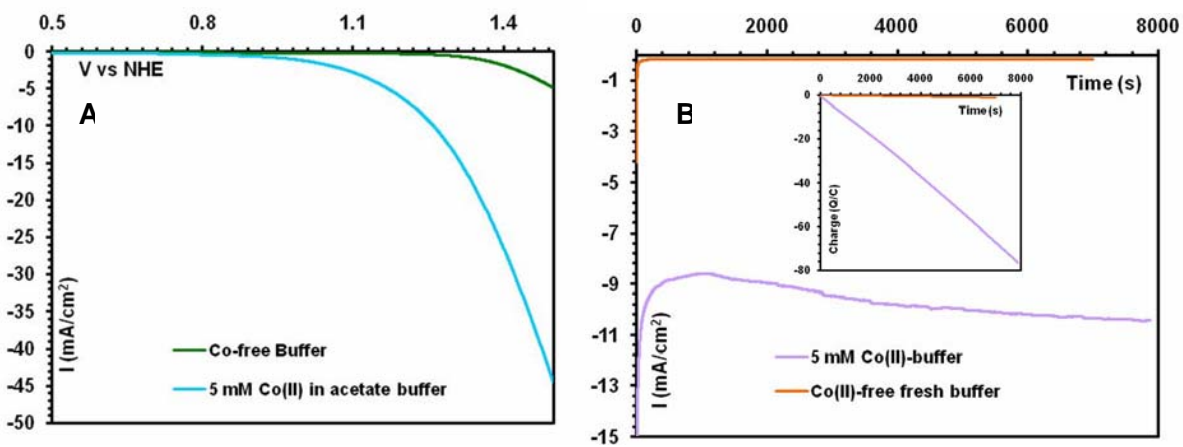

Figure S9: Determination of overpotential (η) by extrapolating the faradic current curve.

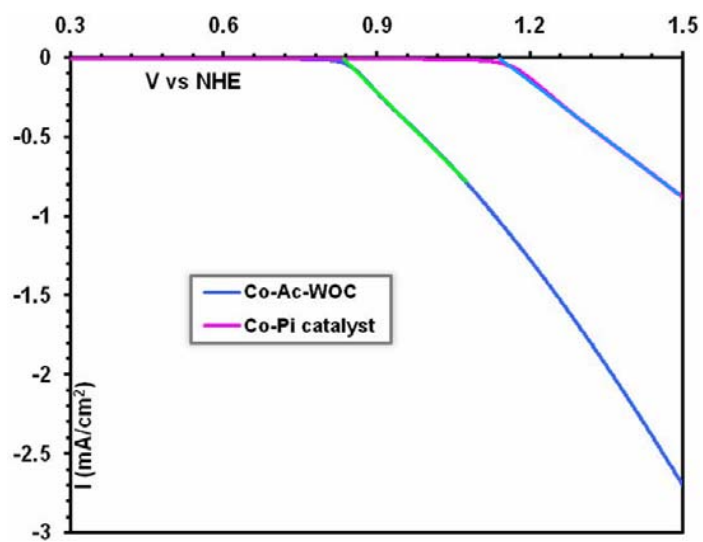

Figure S10: Scan rate (ν) vs current (I) plot. Co-Ac-WOC film (A) on ITO, (B) on FTO.

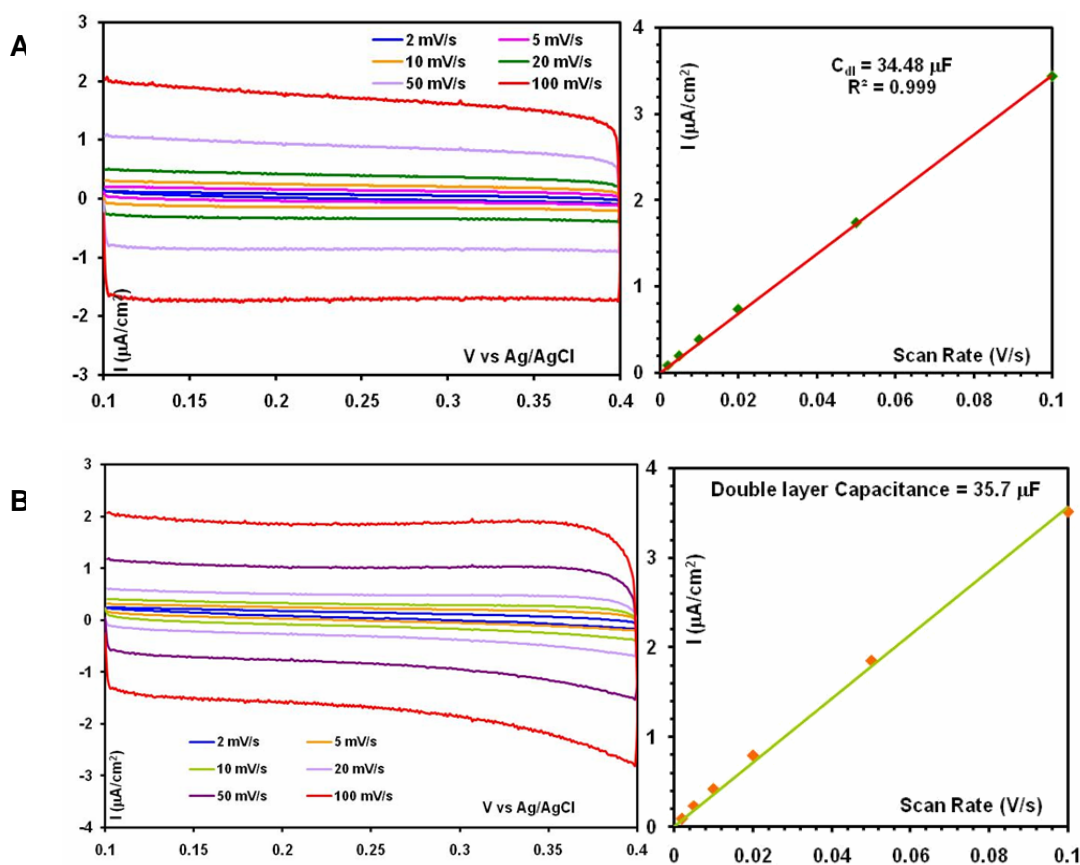

Determination of TOF:

Turn over frequency is determined from the double layer capacitance ($C_{dl}$) of the Co-Ac-WOC film modified ITO or FTO electrode. $C_{dl}$ could easily be determined from the current density of the modified electrode where no faradic current is observed. The difference between the anodic and cathodic current at 0.25 V vs Ag/AgCl (sat. KCl) is plotted against scan rate. From the slope of this curve the double layer capacitance of the modified electrode is easily determined to be ~ 35 µF/cm² for both ITO and FTO electrode. The active surface area for such oxide based material modified electrodes could easily be determined assuming the standard $C_{dl}$ to be 60 µF/cm². Though, this value is very arbitrary, this method is widely acceptable. The TOF has been determined from the double layer capacitance of the Co-Ac-WOC surface ($C_{dl}$ = 35 µF/cm²) and stredy current density (2.5 mA/cm²) which is converted to the number of electrons passed through the electrode per second. The number of electrons could be converted to number of oxygen per second by dividing with 4, as per molecule of oxygen generation involves 4 e⁻ oxidation.

Using these results, the TOF has been determined to be 17.8 s$^{-1}$ at 485 mV overpotentials with ITO electrodes and 7.9 s$^{-1}$ at 385 mV (1.07 mA/cm$^2$) overpotentials with FTO electrodes. This is the highest limiting value as $C_{dl}$ is the measure of the charge passed through the metal centre present at the surface only. So, actual number of metal center participate to the catalysis may be even larger.

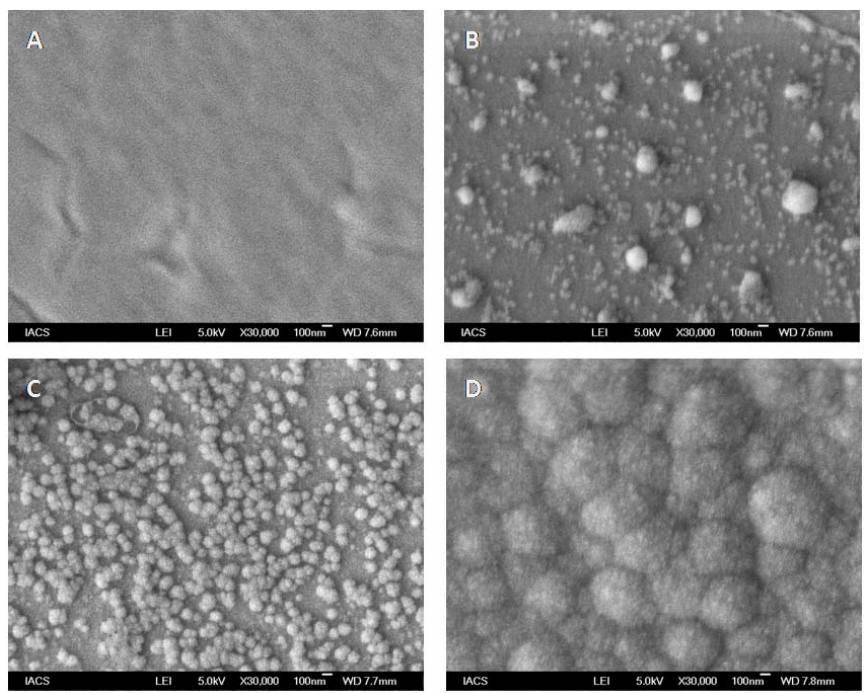

Figure S11: Scanning Electron Microscopy (SEM) of the Co-Ac-WOC modified ITO surface. (A) Electrolysis at 0.75 V vs NHE. (B) Electrolysis at 0.85 V vs NHE. (C) Electrolysis at 1 V. (D) Electrolysis at 1.1 V. All potentials are against NHE.

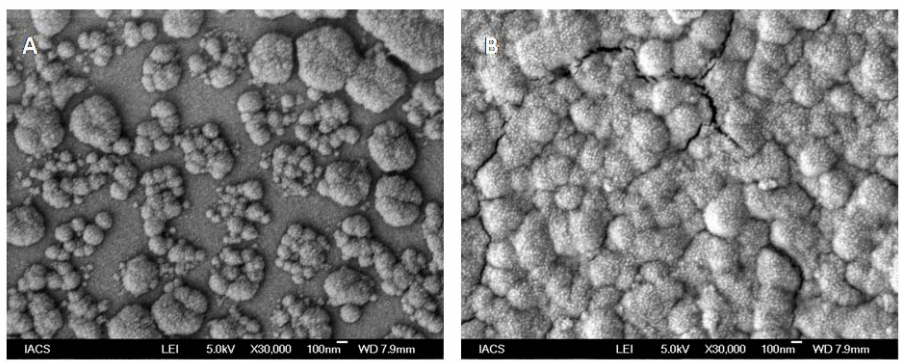

Figure S12: SEM topographic image of the Co-AC-WOC on ITO (A) 5 minutes & (B) 30 minutes.

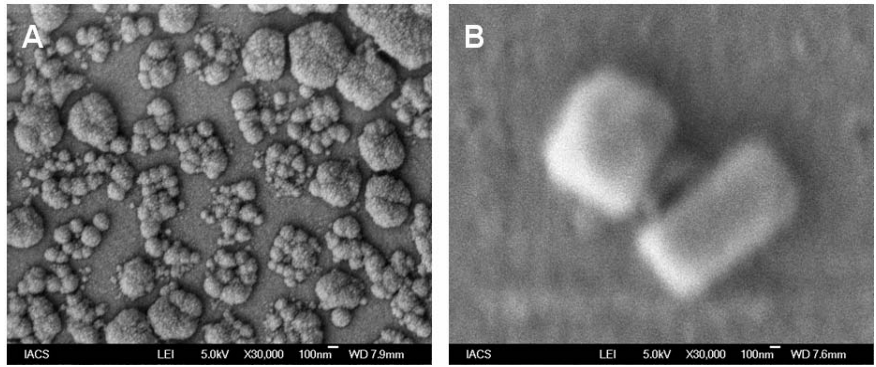

Figure S13: SEM topograph of the modified ITO surface with (A) Co-Ac-WOC and (B) electrodeposited material from Co(II)-salt in potassium acetate buffer with KNO$_3$ electrolyte (pH 7).